\documentstyle[preprint,revtex]{aps}
\math-with-secnums
\begin{document}
\hfill{UM-P-92/88}

\hfill{OZ-92/31}

\begin{title}
\center{The effective potential at finite temperature in the
left-right symmetric model}
\end{title}
\author{J. Choi and R. R. Volkas}
\begin{instit}
Research Centre for High Energy Physics\\
School of Physics, University of Melbourne\\
Parkville, Victoria 3052, Australia
\end{instit}
\begin{abstract}
We obtain the effective potential at finite temperature within the left-right
symmetric model. We take the Higgs sector to consist of two doublets and
calculate its finite temperature effective potential up to the order of
``ring-diagrams'', using analogous techniques to those employed in the
derivation of the effective potential in the Standard Model.
We find that two first order phase transitions can exist,
one at the $SU(2)_R$ breaking scale, the other at the usual $SU(2)_L\otimes
U(1)_Y$ breaking scale. Even though more parameters are present than the
standard model, we find that the electroweak phase transition is still too
weak to allow baryon asymmetry to survive the transition, if the lower
bound of the Higgs boson mass from experiment is adhered to.
\end{abstract}
\newpage

\section{Introduction}

It was Sakharov \cite{oldba} who first pointed out that the observed baryon
asymmetry in the universe may be produced by processes which violate CP
and B, and occur out of thermal equilibrium. One can meet all these
conditions within gauge theories such as the
Standard Model (SM): CP violating terms can be accommodated, sphalerons
can induce sufficient B violating processes at high temperature, and
a first order phase transition can provide the non-thermal equilibrium.
Given this possibility, the electroweak phase transition at high temperature
has been a topic of much interest recently.

To prevent the erasure of any baryon asymmetry produced during the
phase transition, the subsequent sphaleron processes after the phase
transition need to be suppressed by sufficiently high $M_{\rm sph}(T_c)$.
However this condition cannot be met within the minimal SM,
as $M_{\rm sph}(T_c) \propto M_{\rm Higgs}^{-1}$, and the experimental
lower bound of $M_{\rm Higgs}\stackrel{>}{\sim} 57$ GeV places an upper
bound on $M_{\rm sph}(T_c)$ which is too small.

Consequently, there have been attempts to examine the high temperature
behaviour of extensions of the minimal SM. Most notable in this regard
have been the multi-Higgs models and supersymmetric models \cite{turok}
\cite{susy}, which show that one can find models with enough free
parameters to satisfy the sphaleron mass constraint, or conversely
using the constraint to obtain bounds on the parameters.

 Motivated by such approaches, we examine the the finite temperature
effective potential (FTEP) within the left-right symmetric model (LRSM),
with the gauge group of
$\;G_{LR} = SU(2)_L\otimes SU(2)_R\otimes U(1)_{B-L}$.
This model is interesting within the context of anomalous baryon
production for several reasons. First, it is easy to have
sufficient CP violation even with the simplest constructions within
the gauge group \cite{moha1}. Second, it has enough flexibility to allow
for a multitude of possible scenarios that can produce baryon asymmetry.
Third, it has two breaking scales, as $G_{LR}\rightarrow SU(2)_L\otimes
U(1)_Y \rightarrow U(1)_Q$. So we expect to have two distinct phase
transitions. Although the details are unclear, one may expect the
$SU(2)_L$ sphalerons to wash out any baryon asymmetry produced at the
earlier $SU(2)_R$ breaking. In this sense only the $SU(2)_L$ phase
transition seems to have direct relevance to the observed baryon asymmetry.
However we believe the existence of the additional phase transition
scale is interesting in its own right with potential for useful consequences.
In this paper, we will concentrate mainly on writing the finite temperature
effective potential explicitly, by the methods analogous to that used in
the SM. So we begin with a review of the FTEP within the SM \cite{carr}.

\section{The finite temperature effective potential\\ in the standard model}

The SM lagrangian used to calculate the FTEP is usually given by
\begin{equation}
{\cal L}_{SM} = {\cal L}_{gauge field} + {\cal L}_{Higgs} +
                {\cal L}_{fermions},
\end{equation}
where ${\cal L}_{gauge field}$ is the usual sum of
$-\frac{1}{4}F_{\mu\nu}F^{\mu\nu}$ terms,
\begin{equation}
 {\cal L}_{Higgs} = (D_{\mu}\phi)^{\dagger}(D^{\mu}\phi) +
   c^2(\phi^{\dagger}\phi) - \lambda(\phi^{\dagger}\phi)^2
\end{equation} and
\begin{equation}
   {\cal L}_{fermions} = \overline{t_L}\not \!\! D t_L
   + \overline{t_R}\not \!\! D t_R
   + Y\overline{t_L}\phi t_R + {\rm H.c.}
\end{equation}
(neglecting all Yukawa couplings except for the top quark coupling Y).

Following a prescription such as that due to Dolan and Jackiw \cite{dj},
we first shift the field about a classical value $v$; here, we have
\begin{equation}
\phi(x) = \frac{1}{\sqrt{2}}
          \left(\begin{array}{c} v + \phi_1(x) + i\phi_2(x)\\
                                 \phi_3(x) + i\phi_4(x) \end{array} \right).
\end{equation}
The effective potential $V(v)$ is the potential energy of the field $v$ which
is constant in space-time. All masses in the theory which arise via the Higgs
coupling are now functions of the field $v$. Hence the Higgs fields have
masses
\begin{equation}\begin{array}{l}
 m_1^2(v) = 3\lambda v^2 - c^2\\
 m_2^2(v) = m_3^2(v) = m_4^2(v) = \lambda v^2 - c^2
\end{array}\label{SM Higgs mass}\end{equation}
while the gauge bosons and the top quark acquire the masses
\begin{equation}
 m_W^2(v) = g^2v^2/4, \quad m_Z^2(v) = (g^2 + g^{'2})v^2/4,
 \quad m_A^2(v) = 0, \quad m_t^2(v) = Y^2v^2/2.
\label{SM other mass}\end{equation}

The function $V(v)$ is calculated perturbatively using a loop expansion.
To lowest order we have the tree-level potential
\begin{equation}
 V_{tree}(v) = -\frac{1}{2}c^2v^2 + \frac{1}{4}\lambda v^4.
\end{equation}
To next order is the one loop potential with contributions coming from all
the particles in the model:
\begin{equation}
 V_1 (v) = V_{1,\phi}(v) + V_{1,gb}(v) + V_{1,t}(v).
\end{equation}
Usually, the one loop potential is expressed as a sum of temperature
independent and dependent parts:
\begin{equation}
 V_1 (v) = V_1^{(0)}(v) + V_1^{(T)}(v)
\end{equation} where
\begin{equation}
 V_1^{(0)}(v) = \sum_i \frac{n_i}{64\pi^2} m_i^4(v)
                 \left[ ln\frac{m_i^2(v)}{m_{0i}^2} - \frac{3}{2}\right]
\end{equation}
\begin{equation}
 V_1^{(T)}(v) = \sum_i \frac{n_i}{2\pi^2} T^4
                 I_{\pm}\left( \frac{m_i^2(v)}{T^2} \right).
\end{equation}
The sum is over all the particles in the model, with $m_i^2(v)$ as given
in Eqs.~(\ref{SM Higgs mass}) and (\ref{SM other mass}). The quantities
$n_i$ account for the degrees of freedom of each particle,
and $m_{0i}$ is the mass of the particle at zero temperature. The functions
$I_{\pm}$ are given by
\begin{equation}
 I_{\pm}(y) = \int_0^{\infty} dx\,x^2\,ln(1\mp e^{\sqrt{x^2+y}}),
\end{equation}
where $I_+$ is used for bosons and $I_-$ for fermions.

Note that for $v^2<c^2/\lambda\;$, $m_{1,2,3,4}^2(v)$ are all negative,
and the square root in $I_+$ is imaginary. This can be avoided if one
includes the next higher order contribution - ring or ``daisy'' diagrams.
In fact because higher loop graphs have bad infrared behaviour
[there is an effective expansion parameter of order $(T/v)^2$] and the
ring diagrams are the most infrared divergent graphs, taking these into
account can have important consequences \cite{comm}. Calculating the
ring diagrams is essentially a process of resumming perturbation theory
using the self-energies, $\langle\Pi(\omega_n,\bf k)\rangle$.
These give rise to corrections to the zero temperature propagator.
We leave details to Refs~\cite{carr} and state the results here for the SM:
\begin{equation}
 V_{ring,\phi}(v) + V_{1,\phi}^{(T)}(v) =
   \frac{T^4}{2\pi^2}\left[I_+(\frac{m_1^2(v) + \Pi_1(0)}{T^2}) +
        3I_+(\frac{m_2^2(v) + \Pi_2(0)}{T^2}) \right],
\end{equation}
where $\;\Pi_1(0) = \frac{1}{8}g^2T^2 + \frac{1}{16}(g^2 + g^{'2})T^2 +
  \frac{1}{2}\lambda T^2 + \frac{1}{4}Y^2 T^2 = \Pi_2(0)\;$ is the Higgs
self-energy in the infrared limit with contributions coming from
the $SU(2)\otimes U(1)$ gauge bosons, Higgs and top quark fields.
Now the square root in $I_+$ is real for all values of $x$ if
\begin{equation}
 m_i^2(v) + \Pi_i(0) \geq 0, \quad\mbox{or}\quad
 T^2\geq \frac{16c^2}{8\lambda + 3g^2 + g^{'2} + 4Y^2}.
\label{Tmin}\end{equation}

The gauge boson polarization tensors can also be calculated and their
contributions added on to $V_{1,gb}^{(T)}(v)$, even though $I_+$ does not
suffer the negative square root problem here because the gauge boson
masses squared are always positive. The main consequence of taking the gauge
boson polarization tensors into account is, as several authors have noted
recently \cite{comm}, that because the transverse mode of a gauge boson
does not acquire a nonzero $\Pi(0)$ - only the Coulomb mode does -
the strength of the first order phase transition is weakened by a factor
of $2/3$.
This reduces the upper bound on the Higgs boson mass (found by requring
that sphalerons not wash out the baryon asymmetry after the transition) by
a similar factor. As mentioned in the introduction, this makes the minimal
SM unsuitable for baryogenesis through sphalerons.

In summary, the FTEP in the SM may be approximated at high temperature
[$m_i(v)/T < 1$] by using a series expansion for $I_{\pm}$ \cite{AH}\cite{dj}:
\begin{equation}
  V(v) = V_{tree}(v) + V_1^{(0)}(v) + V_{ring,\phi}^{(T)}(v)
          + V_{1,\phi}^{(T)}(v) + V_{1,gb}^{(T)}(v) + V_{1,t}^{(T)}(v)
\end{equation}
where
\begin{equation}\begin{array}{l}
 V_{tree}(v) = -\frac{1}{2}c^2v^2 + \frac{1}{4}\lambda v^4 \\
 V_1^{(0)}(v) = -\frac{3}{2}\frac{1}{64\pi^2}
   \left[m_1^4(v) + 3m_2^4(v) + 3m_Z^4(v) + 6m_W^4(v) - 12m_t^4(v)\right]\\
 \begin{array}{ll}
 V_{ring,\phi}^{(T)}(v)+V_{1,\phi}^{(T)}(v)=\frac{T^2}{24} &
    \left[m_1^2(v) + 3m_2^2(v)\right] \\ & -\frac{T}{12\pi}
    \left[(m_1^2(v)+\Pi_1(0))^{3/2}+3(m_2^2(v)+\Pi_2(0))^{3/2}\right]
 \end{array}\\
 V_{1,gb}^{(T)}(v) = \frac{T^2}{24}\left[3m_Z^2(v) + 6m_W^2(v)\right]
              -\frac{2}{3}\frac{T}{12\pi}\left[3m_Z^3(v) + 6m_W^3(v)\right]\\
 V_{1,\psi}^{(T)}(v) = \frac{T^2}{48}\left[12m_t^2(v)\right].
\end{array}\label{SM pot}\end{equation}
[The $m_i^4(v)ln\frac{m_i^2(v)}{m_{0i}^2(v)}$ terms from $V_1^{(0)}(v)$
get cancelled by corresponding terms in the expansion of
$I_{\pm}(\frac{m_i^2(v)}{T^2})$].
Note that the polarization tensors for the gauge bosons are not
included here but the $\frac{2}{3}$ reduction of the $m_Z^3\;$and$\;m_W^3\;$
terms (which are responsible for the strength of the first order phase
transition) have been explicitly put in. We will use analogous techniques to
those reviewed above to construct the FTEP in the LRSM.

\section{Finite temperature effective potential \\ in the left-right
symmetric model}
We first outline the particular fermion and Higgs spectrum used here.
One commonly used Higgs sector consists of two triplets and a
bi-doublet under $G_{LR}$:
\begin{equation}
 \Delta_L\sim(3,1)(2)\qquad \Delta_R\sim(1,3)(2)\qquad \phi\sim(2,2)(0).
\label{triplets}\end{equation}
Nonzero VEVs for $\Delta_{L,R}$ separate the breaking scale of $G_{LR}$ and
$G_{SM}$, and generate nonzero Majorana masses for the neutrinos.
Phenomenologically the hierarchy
$\langle\Delta_L\rangle\ll\langle\phi\rangle\ll\langle\Delta_R\rangle$ is
necessary, which also leads to the standard see-saw form for the neutrino
mass matrices \cite{trip}. Mohapatra and Zhang \cite{moha2} indicate that
an additional singlet is required to produce the observed baryon asymmetry
within this model. Unfortunately this Higgs sector is rather complicated
for an explicit FTEP construction, as it contains about 15 initial
parameters \cite{gun}.
 Instead, we have turned our attention to the alternative Higgs sector,
with only two Higgs doublets:
\begin{equation}
 \chi_L\sim(2,1)(-1)\qquad \chi_R\sim(1,2)(-1).
\end{equation}
These provide a much simpler Higgs sector than Eq.~(\ref{triplets}).

Now if an exact discrete symmetry $\chi_L\leftrightarrow \chi_R$ exists in
the potential, then either $\langle\chi_L\rangle$ or $\langle\chi_R\rangle$
must be zero if we are to have
$\langle\chi_L\rangle\neq\langle\chi_R\rangle$.
We can avoid this by introducing an additional singlet. However, we consider
the case in which the discrete symmetry $\chi_L\leftrightarrow \chi_R$ does
not hold in the potential \cite{raj}. This not only prevents us from
complicating the potential further with the additional singlet Higgs,
but also enables us to avoid the domain wall problem in cosmology which
stems from spontaneous breaking of a discrete symmetry.

Hence the most general potential is:
\begin{equation}
 V_o(\chi_L, \chi_R) = -\mu_L^2\chi_L^{\dagger}\chi_L
                       -\mu_R^2\chi_R^{\dagger}\chi_R
     +\lambda_L(\chi_L^{\dagger}\chi_L)^2 +\lambda_R(\chi_R^{\dagger}\chi_R)^2
     +2\lambda(\chi_L^{\dagger}\chi_L)(\chi_R^{\dagger}\chi_R),
\label{Vo}\end{equation}
which has only five parameters.
SSB is achieved by
\begin{equation}
 \langle\chi_L\rangle = \frac{\kappa_L}{\sqrt{2}}
                        \left(\begin{array}{c}1\\0\end{array}\right)\qquad
 \langle\chi_R\rangle = \frac{\kappa_R}{\sqrt{2}}
                        \left(\begin{array}{c}1\\0\end{array}\right)\qquad
\end{equation}
where$\qquad \kappa_L^2 = \frac{\lambda_R\mu_L^2-\lambda\mu_R^2}
                               {\lambda_R\lambda_L - \lambda^2}\quad\mbox{and}
       \quad \kappa_R^2 = \frac{\lambda_L\mu_R^2-\lambda\mu_L^2}
                               {\lambda_R\lambda_L - \lambda^2}$.
Clearly one can arrange the parameters such that $\kappa_L\ll\kappa_R$,
hence separating the breaking scales of $G_{LR}$ and $G_{SM}$. We'll
re-express $\mu_L^2$ and $\mu_R^2$ in terms of $\kappa_L^2$ and
$\kappa_R^2$, and use $\kappa_L = 246$ GeV, thus eliminating one parameter.
Positivity of the Higgs mass matrix requires that
$\lambda_L\lambda_R>\lambda^2$.

Because these Higgs doublets cannot couple to any of the existing fermion
bilinears, we need to have exotic fermion multiplets to give masses to
the fermions. The simplest way to do this is by
introduction of vector-like fermions, giving rise to see-saw masses
\cite{raj}. As in the SM, we will assume that only the top quark's Yukawa
coupling is large enough to be important. This implies that the only
fermionic sector of concern to the effective potential is:
\begin{equation}
 {\cal L}_{f} = \overline{t_L}\not\!\!D t_L + \overline{t_R}\not\!\!D t_R
              + [Y_L\overline{t_L}\chi_L T_R +
                Y_R\overline{t_R}\chi_R T_L + {\rm H.c.}]+M_T\overline{T}{T},
\label{Lf}\end{equation}
where $T\sim(1,1)(4/3)$ is the vector-like charge $2/3$ quark.
(We will show later how within the high temperature approximations used,
the other vector-like fermion masses are not parameters of the FTEP even
though they are present to give masses to all the other fermions).

In summary the Lagrangian is given by
\begin{equation}
 {\cal L} = {\cal L}_{gauge field} + {\cal L}_{Higgs} + {\cal L}_f
\end{equation}
where
\begin{equation}
 {\cal L}_{gauge field} = -\frac{1}{4}{\bf F}_{L\mu\nu}{\bf F}_L^{\mu\nu}
                      -\frac{1}{4}{\bf F}_{R\mu\nu}{\bf F}_R^{\mu\nu}
                      -\frac{1}{4}F_{\mu\nu}F_{\mu\nu}
\end{equation}
\begin{equation}
 {\cal L}_{Higgs} =  (D_{\mu}\chi_L)^{\dagger}(D^{\mu}\chi_L)
                   + (D_{\mu}\chi_R)^{\dagger}(D^{\mu}\chi_R)
                   - V_o(\chi_L,\chi_R),
\end{equation}
and $V_o(\chi_L,\chi_R)$ and ${\cal L}_f$ are given by Eqs~(\ref{Vo}) and
(\ref{Lf}) respectively.

Now using a close analogy with the SM case, we construct the effective
potential in the LRSM. First, we shift the two Higgs fields:
\begin{equation}
 \chi_L = \frac{1}{\sqrt{2}}
          \left(\begin{array}{c}v_L+\chi_{1L}+i\chi_{2L}\\
                            \chi_{3L}+i\chi_{4L}\end{array}\right)\qquad
 \chi_R = \frac{1}{\sqrt{2}}
          \left(\begin{array}{c}v_R+\chi_{1R}+i\chi_{2R}\\
                            \chi_{3R}+i\chi_{4R}\end{array}\right),
\end{equation}
from which we obtain $(v_L,v_R)$ dependent masses in the theory:
\begin{equation}\begin{array}{l}
 \begin{array}{ll}
 m_{R,L}^2(v_L,v_R) = \frac{1}{2} &
   \left[ -(\lambda_L+\lambda)\kappa_L^2 - (\lambda_R+\lambda)\kappa_R^2
          +(3\lambda_L+\lambda)v_L^2 + (3\lambda_R+\lambda)v_R^2\right]\\
  &  \pm\sqrt{\frac{1}{4}
     \left[(\lambda-\lambda_L)\kappa_L^2 - (\lambda-\lambda_R)\kappa_R^2
          +(3\lambda_L-\lambda)v_L^2 - (3\lambda_R-\lambda)v_R^2\right]^2
     +4\lambda^2 v_L^2 v_R^2}
 \end{array}\\
 m_{gL}^2(v_L,v_R) = \lambda_L(v_L^2-\kappa_L^2)+\lambda(v_R^2-\kappa_R^2)\\
 m_{gR}^2(v_L,v_R) = \lambda_R(v_R^2-\kappa_R^2)+\lambda(v_L^2-\kappa_L^2)\\
 m_{Z_{R,L}}^2(v_L,v_R) = \frac{g^4}{8(g^2-g_y^2)}
  \left[(v_L^2+v_R^2)\pm\sqrt{(v_L^4+v_R^4)-2(1-2\frac{g_y^4}{g^4})v_L^2v_R^2}
  \right] \\
 m_{W_L}^2(v_L) = \frac{g^2 v_L^2}{4}\\
 m_{W_R}^2(v_R) = \frac{g^2 v_R^2}{4}\\
 m_{T,t}^2(v_L,v_R) = \frac{M_T^2}{2}
   \left[1+\frac{Y_L^2 v_L^2 + Y_R^2 v_R^2}{M_T^2}
        \pm\sqrt{\left(1+\frac{Y_L^2 v_L^2 + Y_R^2 v_R^2}{M_T^2}\right)^2
                 -\left(\frac{2Y_L Y_R}{M_T^2}\right)^2 v_L^2 v_R^2}\right],
\end{array}\label{LR mass}\end{equation}
where $m_{R,L}$ denote Higgs boson masses, $m_{gR,gL}$ denote goldstone
boson masses, and we've used $g = g_L\sim g_R$ as the $SU(2)$ coupling
constants and $g_y$ as the $U(1)_Y$ coupling constant($=g^{'}$ in the SM).
Note that as long as $\kappa_L\neq\kappa_R$,
there is no discrete left-right symmetry in the Lagrangian.
The $g_L\sim g_R$ condition is adopted for the sake of simplicity.

Now we can write down the high temperature expansion of the effective
potential by analogy to Eq.~(\ref{SM pot}):
\begin{equation}\begin{array}{ll}
 V(v_L,v_R) = & V_{tree}(v_L,v_R) + V_1^{(0)}(v_L,v_R)
           + V_{1,\chi}^{(T)}(v_L,v_R)  \\ & + V_{ring,\chi}^{(T)}(v_L,v_R)
           + V_{1,g.b.}^{(T)}(v_L,v_R) + V_{1,\psi}^{(T)}(v_L,v_R),
\end{array}\end{equation}
where
\begin{equation}\begin{array}{l}
 V_{tree}(v_L,v_R) = -\frac{\lambda_L\kappa_L^2 + \lambda\kappa_R^2}{2}v_L^2
                     -\frac{\lambda_R\kappa_R^2 + \lambda\kappa_L^2}{2}v_R^2
         + \frac{\lambda_L}{4}v_L^4 + \frac{\lambda_R}{4}v_R^4
         + \frac{\lambda}{2}v_L^2 v_R^2\\
 \begin{array}{ll}
  V_1^{(0)}(v_L,v_R) = -\frac{3}{128\pi^2}\left[ \right.  &
  m_L^4(v_L,v_R)+3m_{gL}^4(v_L,v_R)+3m_{Z_L}^4(v_L,v_R)+6m_{W_L}^4(v_L,v_R) \\
   & m_R^4(v_L,v_R) + 3m_{gR}^4(v_L,v_R) + 3m_{Z_R}^4(v_L,v_R)
      + 6m_{W_R}^4(v_L,v_R)\\
   &\left. - 12m_t^4(v_L,v_R) - 12m_T^4(v_L,v_R)\right] \end{array}\\
 \begin{array}{l}
  V_{ring,\chi}^{(T)}(v_L,v_R) + V_{1,\chi}^{(T)}(v_L,v_R) = \\
   \qquad\qquad\frac{T^2}{24}
    \left[m_L^2(v_L,v_R) + 3m_{gL}^2(v_L,v_R)
        + m_R^2(v_L,v_R) + 3m_{gR}^2(v_L,v_R)\right] \\
   \qquad\qquad -\frac{T}{12\pi}
    \left[(m_L^2(v_L,v_R)+\Pi_L(0))^{3/2}+3(m_{gL}^2(v_L,v_R)+\Pi_L(0))^{3/2}
    \right]\\
   \qquad\qquad -\frac{T}{12\pi}
    \left[(m_R^2(v_L,v_R)+\Pi_R(0))^{3/2}+3(m_{gR}^2(v_L,v_R)+\Pi_R(0))^{3/2}
    \right]
 \end{array}\\ \begin{array}{ll}
 V_{1,gb}^{(T)}(v_L,v_R) = & \frac{T^2}{24}
         \left[3m_{Z_L}^2(v_L,v_R) + 6m_{W_L}^2(v_L,v_R)
             + 3m_{Z_R}^2(v_L,v_R) + 6m_{W_R}^2(v_L,v_R)\right]\\
      & -\frac{2}{3}\frac{T}{12\pi}
         \left[3m_{Z_L}^3(v_L,v_R) + 6m_{W_L}^3(v_L,v_R)
             + 3m_{Z_R}^3(v_L,v_R) + 6m_{W_R}^3(v_L,v_R) \right]\end{array}\\
 V_{1,\psi}^{(T)}(v_L,v_R) = \frac{T^2}{48}
               \left[12m_t^2(v_L,v_R)+12m_T^2(v_L,v_R)\right]
\end{array}\label{LR pot}\end{equation}
where $\Pi_{L,R}(0)$, the self-energies in the infrared limit, can
be calculated either diagramatically or by
$\Pi_{L,R}(0) = \frac{\partial^2 V_1^{(T)}(v_L,v_R)}{\partial v_{L,R}^2}$.
Keeping only the leading terms, no off-diagonal terms arise and we obtain
\begin{equation}
 \Pi_{L,R}(0) = T^2(\frac{\lambda}{3} + \frac{\lambda_{L,R}}{2} +
          \frac{g^2}{16}\frac{3-2(g_y/g)^2}{1-(g_y/g)^2}+\frac{Y_{L,R}^2}{2}).
\end{equation}
Note that the fermion sector essentially gives the contribution:
\begin{equation}
 V_{1,\psi}^{(T)}(v_L,v_R) = - \frac{T^2}{4}Y^2(v_L^2+v_R^2),
\end{equation}
dropping $v_L,\;v_R$ independent terms and where $Y=Y_L\sim Y_R$ for
simplicity. Hence even if we were to include the effects of the other
fermions, such as the electron and its vector-like partner $E$, their
contributions will be $-\frac{T^2}{4}Y_E^2(v_L^2+v_R^2)$. Hence as long
as $Y^2\gg Y_E^2$ (large top Yukawa coupling), their effects are negligible
to the FTEP \cite{dof}.

What can we expect from this potential? If the SSB pattern of the gauge
groups is to be realised as the temperature is varied, one expects two phase
transitions: one at $T=T_R\sim O(\kappa_R) (\stackrel{>}{\sim} 10 TeV)$
where $SU(2)_R$ is spontaneously broken, and the other at
$T=T_L\sim O(\kappa_L) (= 246 GeV)$. Hence at temperatures much higher than
$T_R$ down to $T=T_R$, $v_L=v_R=0$ will be the minimum of the FTEP.
At $T=T_R$, two degenerate minima exist: one for $v_L = v_R = 0$, and
a new one at $v_L=0$, $v_R=V_R(T_R)$. For temperatures $T<T_R$, the
right sector Higgs field $v_R$ will settle down near the minimum
given by $v_R=V_R(T)$. The quantity $V_R(T)$ will then slowly evolve
with $T$ to its zero-temperature value $\kappa_R$. Of course, in the
evolution of the universe, the dynamics of the right-sector phase
transition will be complicated. For instance, if the transition is
first order, then an out-of-equilibrium situation will persist for
some time, and so the use of the FTEP calculated within equilibrium
thermodynamics will be of little use. However, we will assume that
by the time the usual left-sector phase transition occurs, equilibrium
has again been attained so that our FTEP can again be reliably used,
as long as we restrict our attention to $V_R\simeq \kappa_R$ region.
We can represent the expected behaviour of the FTEP as follows:
\begin{equation}\begin{array}{lcc}
 T \qquad\quad & \mbox{min. of}\; V(v_L,v_R)\; \mbox{occurs at} &
      \qquad\quad\mbox{gauge symmetry}\\ \hline
 \gg T_R & (0,0) & G_{LR} \\
 T_R & (0,0)\;\mbox{and}\;(0,V_R(T_R)) & G_{L,R}\rightarrow G_{SM}\\
 <T_R & (0,V_R(T)) & G_{SM}\\
 T_L & (0,\kappa_R)\;\mbox{and}\;(V_L(T_L),\kappa_R) &
       G_{SM}\rightarrow U(1)_Q\\
 <T_L & (V_L(T),\kappa_R) & U(1)_Q\\
 0 & (\kappa_L, \kappa_R) & U(1)_Q
\end{array}\label{table}\end{equation}

Note that all the masses [Eq.~(\ref{LR mass})] are also separated by a
large scale in the regions of $(v_L,v_R)$ space of interest - namely
the values of $m_{iR}^2$ far exceed those of $m_{iL}^2$ near the regions
$(0,V_R(T_R))$ and $(V_L(T_L),\kappa_R)$. This means that near $T_L$,
where all $m_{iR}^2$ are large, we cannot use the high temperature
expansion of $I_{\pm}$, Eq.~(\ref{LR pot}). However since $m_{iR}\gg T$
there, the contributions due to these particles will be Boltzmann suppressed,
and we can add on their contribution by
\begin{equation}
 V_T(v_L,\kappa_R) = -\sum_i \frac{n_i T^2}{(2\pi)^{3/2}}
   m_{iR}^2(v_L,\kappa_R) \sqrt{\frac{T}{m_{iR}}}e^{-m_{iR}/T}
   \left[1+\frac{15}{8}\frac{T}{m_{iR}} +
   O\left(\frac{T^2}{m_{iR}^2}\right) \right].
\label{Boltz}\end{equation}
Hence the effective potential near $T_L$ is given by Eq.~(\ref{LR pot}) with
all $m_{iR}^2$ terms removed, plus Eq.~(\ref{Boltz}).
Alternatively, using numerical integration, we can use the exact expressions
of $I_{\pm}$ which will be valid for all $T$. In any case we find that near
$(0,V_R)$ at $T\sim T_R$, all $m_{iL}^2$ terms give negligible contributions
compared to $m_{iR}^2$ terms, while near $(V_L,\kappa_R)$, $m_{iR}^2$ terms
give negligible contributions. Hence the model resembles writing two FTEPs
separated by a large scale. There are two minimum temperatures required to
keep the square root in $I_{\pm}$ real [the equivalent of Eq.~(\ref{Tmin})]:
one near $T_L$ where $v_R\simeq\kappa_R$ and another near $T_R$ where
$v_L\simeq 0$. These are given respectively by:
\begin{equation}\begin{array}{ll}
 T_{Lmin}^2 = {\rm Max}[ & \frac{\lambda_L\kappa_L^2}{G_L},\;
                          \frac{\lambda\,\kappa_L^2}{G_R}]\\
 T_{Rmin}^2 = {\rm Max}[ & \frac{\eta_L^{+} + \eta_R^{+}
                             + \left| \eta_R^{-}-\eta_L^{-}\right|}{2G_L},
                           \frac{\eta_L^{+} + \eta_R^{+}
                             - \left| \eta_R^{-}-\eta_L^{-}\right|}{2G_R},
                       \frac{\lambda\,\kappa_R^2 + \lambda_L\kappa_L^2}{G_L},
                       \frac{\lambda\,\kappa_L^2 + \lambda_R\kappa_R^2}{G_R}],
\end{array}\end{equation}
where $\eta_{L,R}^{\pm}=(\lambda_{L,R}\pm\lambda)\kappa_{L,R}^2$ and
$G_{L,R}$ are defined by $\Pi_{L,R}(0)=T^2 G_{L,R}$.

\section{Discussion}
We have used the following parameters for definiteness:
\begin{equation}\begin{array}{c}
 \kappa_L=246\,{\rm GeV},\qquad g_R\sim g_L = g = 0.652,\qquad g_y = 0.352 \\
 M({\rm top})=125\,{\rm GeV},\qquad M(W_R)=5\,{\rm TeV} \;\;
  (\Rightarrow \kappa_R\sim 16\,{\rm TeV})
\end{array}\end{equation}
leaving us with 3 tunable parameters $\lambda_L, \lambda_R$ and $\lambda$.
We are first interested in seeing whether or not the $V(v_L,v_R)$ has
the qualitative features described in Eq.~(\ref{table}).
Fig~(\ref{vlow1}) shows a typical behaviour of the potential near $T_L$,
while Fig~(\ref{vhigh1}) shows this near $T_R$. Table~(\ref{params})
shows the relationship among transition temperature, Higgs masses and
the values of VEV where first order phase transitions occur, for some
values of chosen $\lambda$'s. We have used the high $T$ expansion of
$I_{\pm}$ to obtain these results, checking that the parameters chosen are
such that $m_i<T$ near the regions shown. Also note that perturbation
theory will break down for $v_{L,R}$ close to zero. Typically this
will happen for $v_{L,R}/T \stackrel{<}{\sim} 0.1$ \cite{carr}
and we are at best extrapolating here.

Next, we want to see if the extra parameters in this model allow
for large $M_{\rm sph}$ after the phase transition while at the same time
allowing $M_{\rm Higgs}\stackrel{>}{\sim} 57$ GeV. The mass of the
sphaleron within the SM is given by
$M_{\rm sph}(T) =\frac{4\pi}{g}B(\lambda)\phi(T)$ \cite{klink},
where $1.56<B(\lambda)<2.72$. The constraint that sphaleron processes are
suppressed after a phase transition is given by
$\frac{M_{\rm sph}(T_c)}{T_c}\geq 45$ \cite{shapo} for some transition
temperature $T_c$. If we assume that we can use the SM expression of
$M_{\rm sph}(T)$ at the electroweak phase transition temperature, $T_L$,
(this assumption appears reasonable since the LRSM approximates the SM
well below $T_R$) then we want
\begin{equation}
 \frac{V_L}{T_L}\stackrel{>}{\sim}1.46 \qquad\qquad \mbox{using the maximum }
 B(\lambda),
\label{sph condn}\end{equation}
in order to suppress the sphaleron rate \cite{mdept}.

Although we have more free parameters here than in the SM, we find that
Eq~(\ref{sph condn}) still cannot be satisfied. This is due to to the
smaller Higgs mass at zero temperature, $m_L(\kappa_L,\kappa_R)$, and
the ratio $V_L/T_L$ being mostly sensitive to $\lambda_L$, rather than
$\lambda_R$ or $\lambda$. We still find that smaller $\lambda_L$ favours
larger $V_L/T_L$ but produces smaller $m_L(\kappa_L,\kappa_R)$ at the same
time. Table~(\ref{params}) shows this (we chose some of the parameters
with a bias towards producing a large $V_L/T_L$ while keeping the Higgs
mass large). Although we have used some parameter simplifications by
setting $g_L=g_R$ and $Y_L=Y_R$, as well as using fixed values for
$\kappa_R$ and $M_{\rm top}$, we suspect the ratio $V_L/T_L$ will still
not change significantly if these are varied (it may well improve the
ratio $V_R/T_R$, but this will not be relevant to the survival of baryon
asymmetry, as mentioned in Sec~III) \cite{gkap}. However one would expect
to overcome this in simple extensions of this model. If a singlet is added
on to the Higgs sector for instance, a cubic term would be generated at tree
level, making the phase transition more strongly first order at
$T_L$ \cite{piet}.

\section{conclusion}
We have calculated the effective potential at finite temperature within
the left-right symmetric model with two Higgs doublets. We find that the
effective potential can have two first order phase transitions, one at
the $SU(2)_R$ breaking scale and the other at the usual electroweak breaking
scale. However it is impossible to simultaneously have baryon asymmetry
surviving a first order electroweak phase transition
and an electroweak Higgs boson mass greater than $\sim 57$ GeV.
Therefore the qualitative conclusions regarding electroweak baryogenesis
arising from the simplest possible Higgs sector of the left-right symmetric
model are the same as for the minimal standard model. However, we can
now build on the initial study of the finite-temperature effective potential
within the left-right symmetric model performed in this paper.
We expect that more complicated Higgs sectors (eg. $\chi_L,$ $\chi_R$
plus a gauge singlet Higgs field) will produce more promising scenarios
for low-temperature baryogenesis in
$SU(2)_L\otimes SU(2)_R\otimes U(1)_{B-L}$ electroweak models. We hope
to return to this issue in the future.

\section{acknowledgments}
The authors would like to thank N. Cottingham for discussions and H. Lew
for reading the manuscript and for comments. One of
us (J.C.) would like to acknowledge the support of the Australian
Postgraduate Research Program and thank the following people for
useful discussions: C. Dettmann, J. Daicic, A. Davies, P. W. Dyson,
B. Hanlon, L. Hollenberg, B. Smith and A. Waldron. R. R. V. is
supported by the Australian Research Council through a Queen Elizabeth
II Fellowship.

\figure{The cross-section of the effective potential at $v_R=\kappa_R$,
 near $T_L$. The Higgs boson masses are taken to be $69.2$ and  $4338$ GeV.
 \label{vlow1}}
\figure{The cross-section of the effective potential at $v_L=0$,
 near $T_R$. The Higgs boson masses are taken to be $69.2$ and  $4338$ GeV.
 \label{vhigh1}}

\begin{table}
\caption{First order phase transition temperatures and the Higgs
masses (all in GeV) for some values of the parameters. To take the
constraint $\lambda_L\lambda_R>\lambda^2$ into account, we have
defined $\lambda=(\lambda_L\lambda_R)^{1/2}\Delta\lambda$, so that
$0<\Delta\lambda<1$. $V_{L,R}$ are the values of $v_{L,R}$ at which
degenerate minima occur, and $m_H$ are the two Higgs boson masses at
zero temperature.}
\[\begin{array}{cccccccc}
 \lambda_L\quad & \lambda_R\quad & \Delta\lambda\quad & m_H\quad
            & T_L\quad & V_L\quad & T_R\quad & V_R \\ \hline
 0.028\quad & 0.002\quad & 0.07\quad & 58.1, 970\quad & 87.26\quad
            & 33\quad & 1071.5\quad & 310\\
 0.04\quad & 0.002\quad & 0.07\quad & 69.4, 970\quad & 103.33\quad
            & 30\quad & 1071.5\quad & 310\\
 0.1\quad & 0.002\quad & 0.07\quad & 109.7, 970\quad & 157.15\quad
            & 21\quad & 1071.5\quad & 300\\
 0.028\quad & 0.002\quad & 10^{-3}\quad & 58.2, 970\quad & 87.26\quad
            & 33\quad & 1071.5\quad & 310\\
 0.028\quad & 0.002\quad & 10^{-4}\quad & 58.2, 970\quad & 87.26\quad
            & 33\quad & 1071.5\quad & 310\\
 0.1\quad & 0.005\quad & 0.045\quad & 110, 1534\quad & 157.15\quad
            & 22\quad & 1691.2\quad & 460\\
 0.1\quad & 0.01\quad & 0.28\quad & 106, 2170\quad & 157.49\quad
            & 21\quad & 2378.4\quad & 600\\
 0.1\quad & 0.1\quad & 0.5\quad & 95, 6859\quad & 158.26\quad
            & 21\quad & 7099.8\quad & 950

\end{array}\]
\label{params}
\end{table}

\begin{references}
\bibitem{mdept} Although the experimental lower bound of light Higgs
                will be model dependent, our Higgs and fermion sectors
                should not modify the results significantly.
\bibitem{dof}   Note that the bare masses for the vector-like fermions
                are independent of $v_L$ and $v_R$.
\bibitem{gkap}  Although we have not performed an exhaustive search of
                $(g_R, \kappa_R)$ parameter space, we are confident that
                our qualitative conclusion regarding sphaleron
                wash-out of the baryon asymmetry after the left-handed
                breaking scale is correct. We performed a coarse search
                through this parameter space, and found no promising trends.
                Therefore, given our previous physical argument that
                the most important parameter for this issue is $\lambda_L$,
                our conclusion seems to be sound.
\end{references}
\end{document}